# Silicon graphene waveguide tunable broadband microwave photonics phase shifter

**José Capmany[1],* David Domenech[2], and Pascual Muñoz[1]**

[1]*ITEAM Research Institute, Universitat Politécnica de Valencia, Camino de Vera s/n, 46022 Valencia, Spain*
[2]*VLC Photonics S.L., Camino de Vera s/n, 46022 Valencia, Spain*
* *jcapmany@iteam.upv.es*

**Abstract:** We propose the use of silicon graphene waveguides to implement a tunable broadband microwave photonics phase shifte based on integrated ring cavities. Numerical computation results show the feasibility for broadband operation over 40 GHz bandwidth and full 360º radiofrequency phase-shift with a modest voltage excursion of 0.12 volt.

## 1. Introduction

Graphene is a two dimensional single layer of carbon atoms arranged in an hexagonal (honeycomb) lattice that has generated considerable interest in recent years due to its remarkable optical and electronic properties [1]-[3]. It has an energy vs momentum dispersion diagram where the conduction and valence bands meet at single points (Dirac points) [1]. In the vicinity of a Dirac point, the band dispersion is linear and electrons behave as fermions with zero mass, propagating at a speed of light of around $10^6$ m $s^{-1}$ and featuring mobility values of up to $10^6$ $cm^2V^{-1}s^{-1}$. These electronic properties make this material a potential candidate for nanoelectronics and, in particular, for high-frequency applications [1]-[3]. Graphene also shows unusual optical properties [4]. For instance, due to its linear dispersion, it can absorb light over a broad frequency range enabling broadband applications. Saturable absorption can be observed as a consequence of Pauli blocking, and non-equilibrium carriers result in hot luminescence.

Another remarkable property of graphene is that the density of states of carriers near the Dirac point is low, and as a consequence, its Fermi energy can be tuned significantly with relatively low electrical energy (applied voltage) [1]-[4]. This Fermi level tuning changes, in turn, the refractive index of graphene. Thus, the combination of graphene with integrated dielectric waveguides opens unprecedented possibilities for the design of tunable components and several groups have recently reported devices with applications in the microwave [5], terahertz [3] and photonic regions [4] of the electromagnetic spectrum. A particularly active area of research during the last years is related to the design of tunable integrated photonic components where different contributions have theoretically and experimentally reported a variety of functionalities including electroabsorption modulation in straight waveguides [5]-[7] and resonant structures [8], channel switching [9], and electrorefractive modulation [10].

Most of the former components can find application in digital communications but, to our knowledge, graphene has not been yet proposed for the implementation of tunable analog photonic components [12], [13] and, in particular, for devices providing essential functionalities in the emerging field of integrated microwave photonics [14], such as broadband phase shifters and true time delay lines. In this paper we propose the use of graphene in combination with silicon waveguide resonators for the implementation of integrated broadband fast tunable microwave photonic phase shifters. The paper is structured as follows. In section 2 we review some basic properties of graphene that are important in the design of tunable photonic components, most notably the dependence of its conductivity and dielectric constant on the chemical potential and applied voltage. In section 3 we propose a silicon photonics waveguide design for the implementation of the phase shifter and evaluate the impact that the incorporation of the graphene layer has on the effective index and absorption of its fundamental TM and TE guided modes. We pay special attention to the impact that voltage change has on the tunability of both parameters and identify the suitable bias regions for electroabsorption and electrorefractive operation. In section IV we present the integrated phase shifter design. It is based on single-sideband modulation of the broadband RF signal onto an optical carrier that is then shifted relative to the modulated signal using the resonance of an optical all-pass ring cavity filter. This phase shifting can be tuned by incorporating graphene into the waveguide implementing the ring cavity. Broadband operation in the 10-50 GHz region is demonstrated and the design is expanded to a two stage cavity in order to provide full 360º range phase shift. Section V presents some conclusions and future directions.

## 2. Graphene conductivity and dielectric constant

Graphene has noteworthy optical properties due to its conical band structure that allow both intraband and interband transitions [1]-[3]. Both types of transitions contribute to the material conductivity [15].

$$\sigma(\omega) = \sigma_{\text{intra}}(\omega) + \sigma_{\text{inter}}(\omega) \tag{1}$$

For instance, intra-band transitions are the dominant source for the overall conductivity in the microwave and terahertz regions of the spectrum that can be expressed in terms of the Kubo's formula [15]:

$$\sigma_{\text{intra}}(\omega) = \frac{ie^2}{\pi\hbar^2(\omega + i2\Gamma)}\left[\frac{\mu_c}{k_B T} + 2\ln\left(e^{-(\mu_c/k_B T)} + 1\right)\right] \tag{2}$$

Where $e$ represents the charge of the electron, $\hbar$ the angular Planck constant, $k_B$ the Boltzman constant, $T$ the temperature, $\mu_c$ is the Fermi level or chemical potential and:

$$\Gamma = \frac{e\mathbf{v}_\mathbf{F}^2}{\mu\mu_c} \tag{3}$$

is the electron collision rate which is a function of the electron mobility $\mu$ and the Fermi velocity in graphene $\mathbf{v}_\mathbf{F} \approx 10^6 ms^{-1}$. In the visible optical region of the spectrum however, inter-band transitions dominate the conductivity that is given if $k_B T << |\mu_c|, \hbar\omega$ by [15]:

$$\sigma_{\text{inter}}(\omega) \approx \frac{-ie^2}{4\pi\hbar}\ln\left(\frac{2|\mu_c| - (\omega - 2i\Gamma)\hbar}{2|\mu_c| + (\omega - 2i\Gamma)\hbar}\right) \tag{4}$$

Form (1)-(4) one can get the dielectric constant of a layer of graphene:

$$\varepsilon_g(\omega) = 1 + \frac{i\sigma(\omega)}{\omega\varepsilon_o\Delta} \tag{5}$$

Where *Δ=0.34 nm* is the thickness of the layer. The left part of Figure 1 represents as an example, the real and imaginary parts of both the intra and inter-band conductivities for $\lambda = 1550nm$, $T = 300^o K$, $1/2\Gamma = 5.10^{-13}$ sec as a function of the chemical potential, while the lower part represents the dielectric constant. In this particular example a transition can be observed at $|\mu_c| = 0.4eV$, where the dielectric constant changes for purely real $|\mu_c| > 0.4eV$ to imaginary $|\mu_c| < 0.4eV$. Note that in the vicinity of $|\mu_c| \geq 0.4eV$ a small change in the chemical potential yields a substantial change in the real value of the effective index of graphene as shown in the right part of figure 1. Graphene is electrorefractive in that region. On the other side, a small change in the chemical potential in both directions around $|\mu_c| = 0.4eV$ yields a substantial change in the imaginary value of the dielectric constant (i.e the losses) and graphene is electroabsorptive in that region.

Exploiting either the electrorefractive behaviour of graphene lies at the heart of designing the microwave photonics phase shifter. Tunability is achieved by suitable application of a voltage $V_g$ to the graphene layer, since this changes the value of the chemical potential according to [11]:

$$\left|\mu_c\left(V_g\right)\right| = \hbar \mathbf{v_F}\sqrt{\pi\left|\eta\left(V_g - V_o\right)\right|} \qquad (6)$$

Where *Vo=0.8* volt is the offset from zero caused by natural doping and $\eta = 9\text{x}10^{16} V^{-1} m^{-2}$ [6].

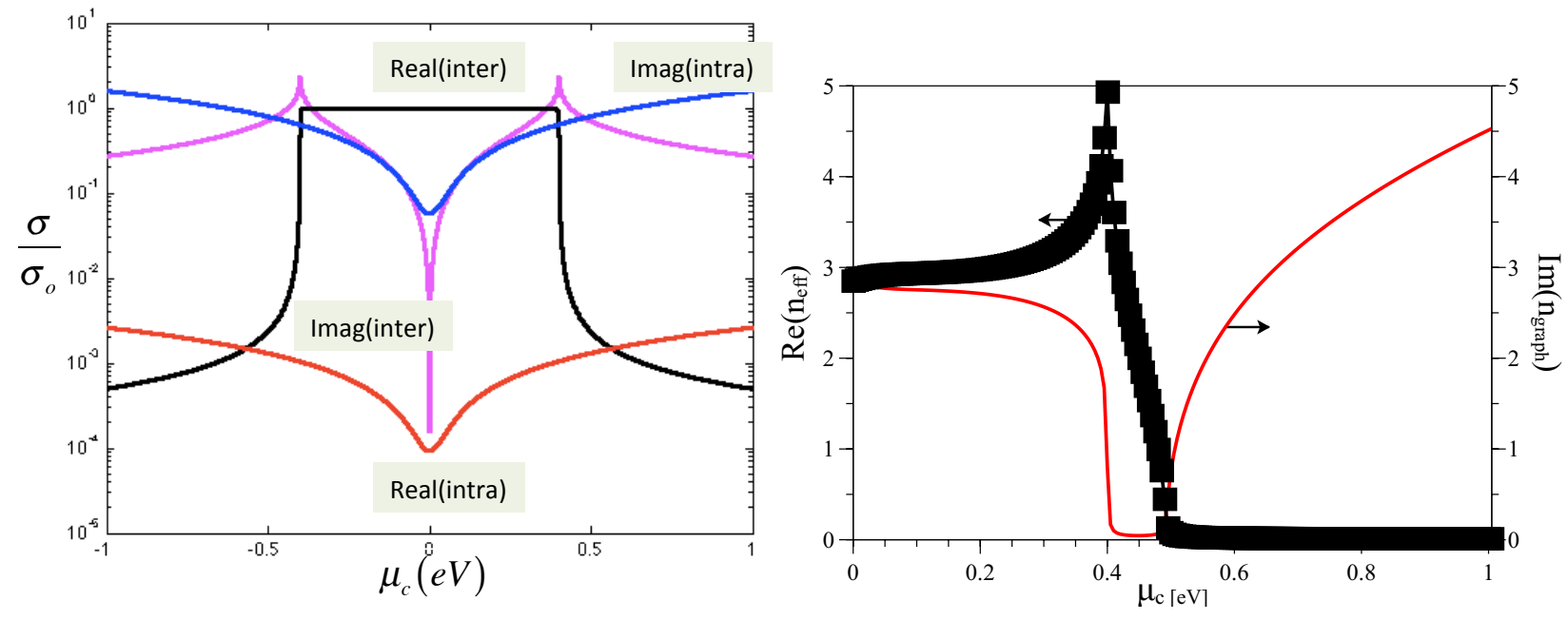

Fig. 1. (Left) Real and imaginary parts of the intra-and inter-band conductivities of graphene for $\lambda = 1550nm$, $T = 300^o K$, $1/2\Gamma = 5.10^{-13}$ sec . (Right) Overall complex effective index.

## 3 Graphene silicon waveguide

Graphene can be incorporated into silicon to implement graphene silicon waveguides (GSWs). One approach, shown in figure 2 consists in placing a monolayer graphene sheet on top of a silicon bus waveguide, separated from it by a thin $Al_2O_3$ layer.

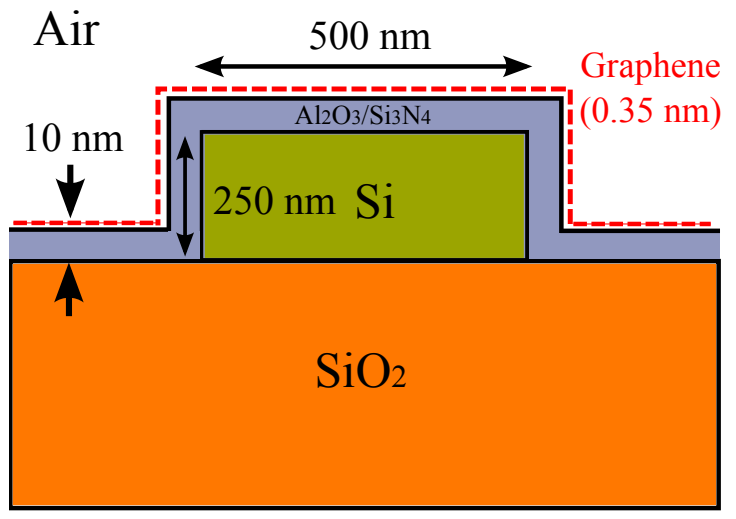

Fig. 2. Deep silicon waveguide with a layer of graphene placed on top of it.

The presence of the graphene layer modifies the propagation characteristics (field profile, losses and effective index) of the guided modes and these can be in turn, as mentioned above, controlled and reconfigured changing the chemical potential by means of applying a suitable voltage. In addition, all these properties are wavelength dependent so a complete description of how these parameters change in terms of chemical potential and wavelength is required.

With the exception of very simple and unpractical waveguide configurations, this description requires the use of numerical and or mode solving techniques.

The transverse electric (TE) and Transverse Magnetic (TM) modes of the waveguide have been numerically calculated by means of a Finite Differences (FD) based commercial *Field Designer* mode solver, from *PhoeniX Software B.V.* The dispersive nature of the materials has been taken into account in the numerical calculations in the wavelength range under study (1540-1560 nm). The left part of figure 3 represents, for λ=1550 nm, the effective index and the losses (in insets) vs the chemical potential for both the TE and TM fundamental modes. The right part provides as well the wavelength dependence.

Note that the main electrorefractive and electroabsorptive characteristics of graphene displayed in figure 1 are transferred to the waveguide TE and TM modes, despite the small dimensions of its layer as compared to those of the waveguide. This fact and the subsequent possibility of changing the effective index the guided modes with the applied voltage opens the possibility of controlling the phase of the signals propagating through these waveguides. In the next section we apply this effect to the design of the microwave photonics phase-shifter.

## 4 Tunable microwave photonic phase shifter

A tunable broadband microwave phase shifter is a key component in applications such as phased antenna arrays, microwave filters and reconfigurable front-ends [12],[13]. A particularly simple and versatile design is that based on single-sideband modulation of the broadband RF signal onto an optical carrier [16]. The phase of the carrier is then shifted relative to the modulated signal using the resonance of an optical all-pass ring cavity filter as shown in the upper part of figure 4.

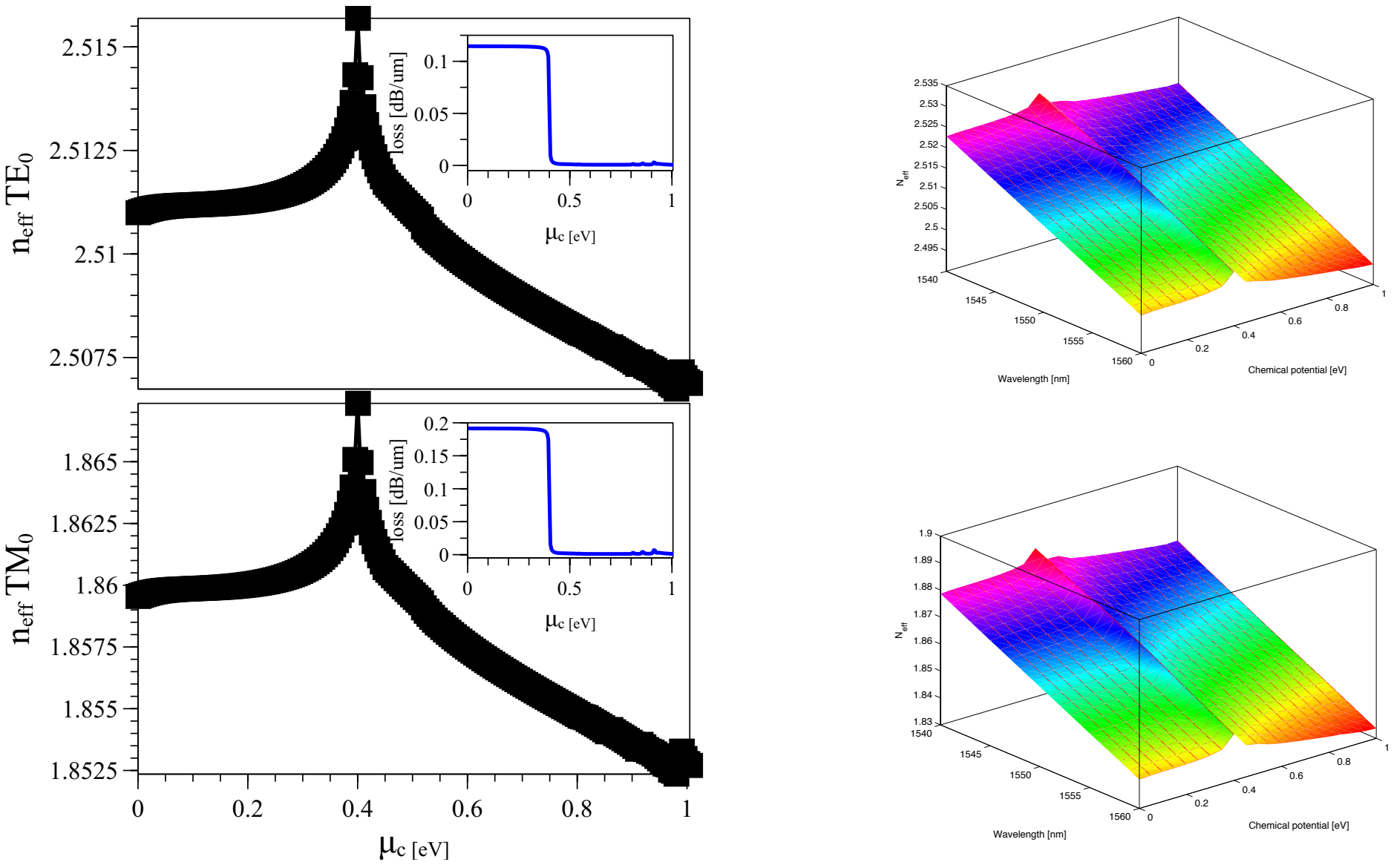

Fig. 3. (Left) Effective indices and losses for the TE (upper) and TM (lower) modes of a deep GSW. (Right) 2D Effective indices and losses for the TE (upper) and TM (lower) modes of a deep GSW versus the wavelength and the chemical potential.

Integrated versions of this phase shifter have been reported in silicon on Insulator (SOI) [17] and Silicon Nitride TriplexX technologies [18]. In both cases however the tuning mechanism is based on thermal effects and thus is slow-speed. The high electron mobility of graphene suggests that a voltage-controlled fast tunable phase shifter could be obtained by incorporating graphene into the ring waveguide. The voltage control signal would act on the chemical potential, thus changing the resonance position of the filter. The lower part of figure 4 illustrates, as an example, the design of a TM phase-shifter based on the deep GSW reported in section 3.

The structure is composed of two cavities rather than a single one in order to easily achieve a sharper transition in the phase transfer around the resonance frequency. The structure is designed for operation at 1550 nm featuring a ring radius is 250 μm which yields a free spectral range of 100 GHz, enough for broadband operation of the phase shifter in the 10-50 GHz Rf frequency range. The resonator coupling constants are both equal to 0.12.

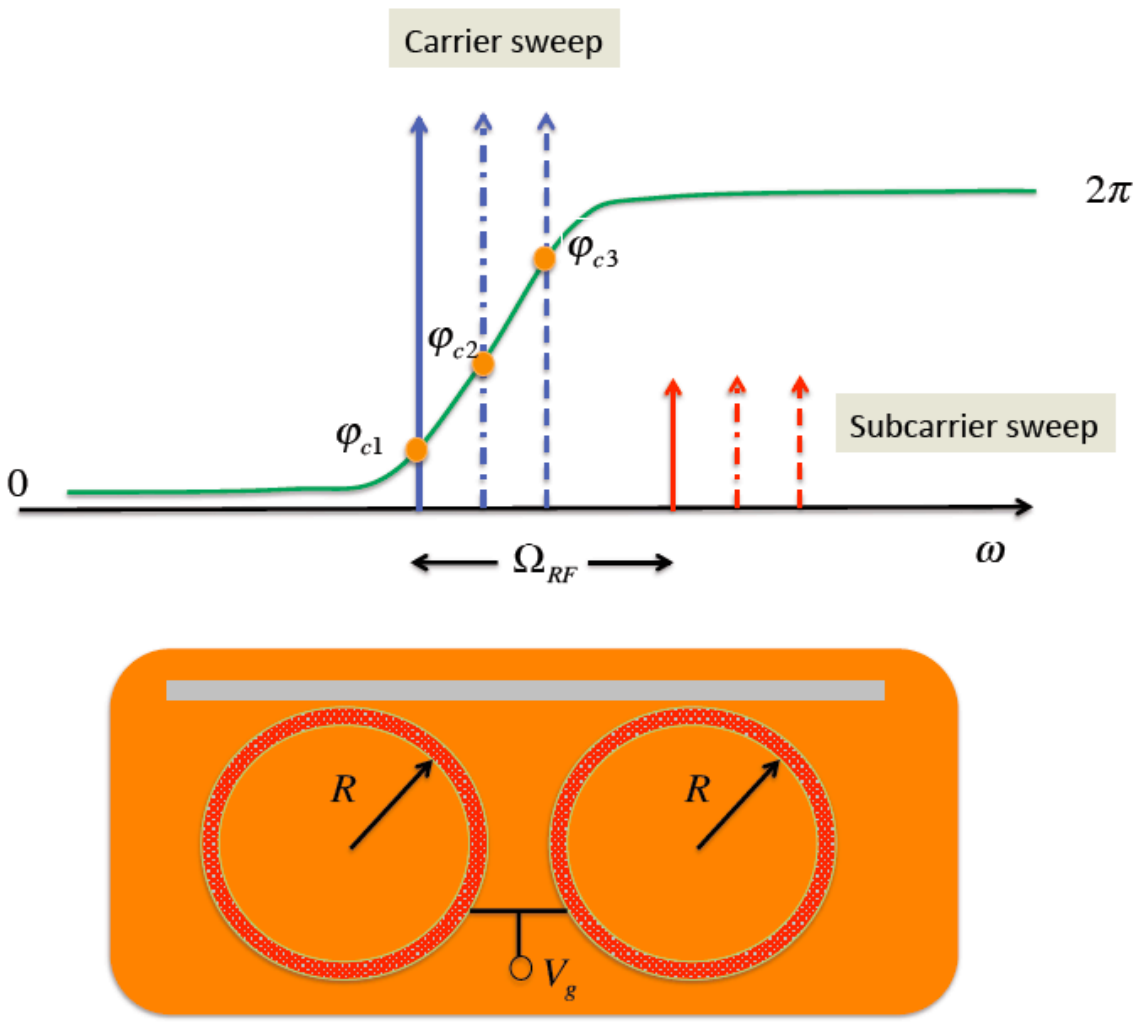

Fig.4. (Upper) principle of operation of a single-sideband microwave photonics RF phase shifter using an all pass optical filter resonance. (Lower) two stage RR GSW design of a tunable phase shifter (parameter details are given in the text).

Figure 5 displays the results for the designed phase shifter. The left part represents the impressed phased shift over the detected RF subcarrier (modulo 2π) as a function of the chemical potential of graphene where the RF frequency is taken as a parameter. Note that there is a region of the chemical potential in between 0.5656 and 0.567 eV where a semilinear phase-shift vs chemical potential spanning 360º is obtained.

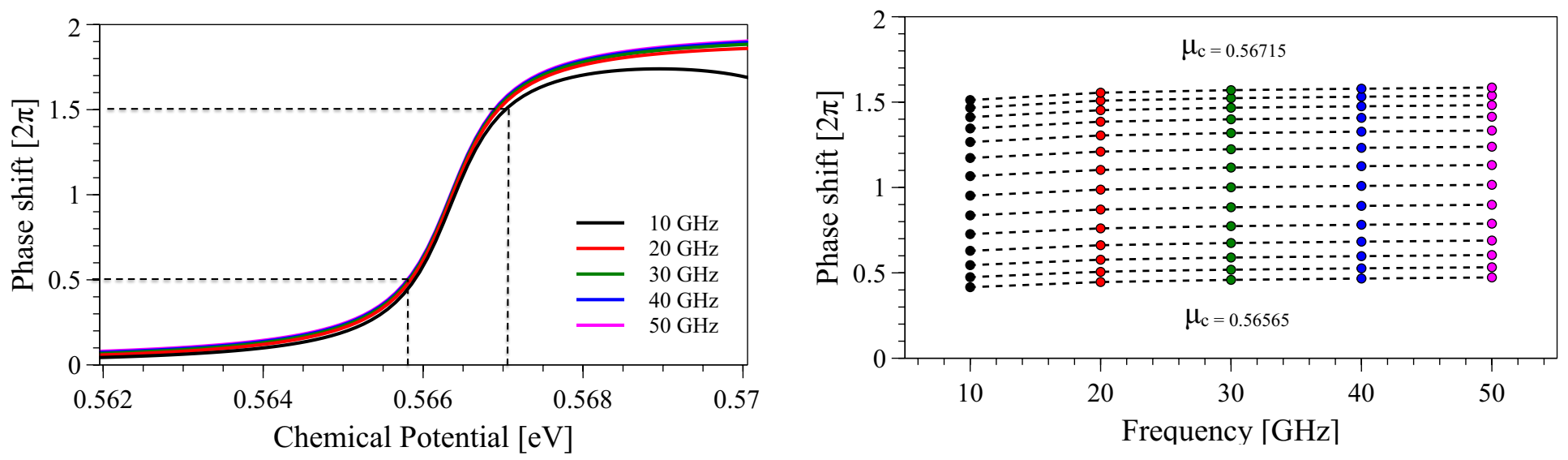

Fig. 5. (Left) impressed phased shift over the detected RF subcarrier (modulo 2π) as a function of the chemical potential of graphene where the RF frequency is taken as a parameter. (Right) impressed phase shift versus the subcarrier frequency, taking the chemical potential as a parameter.

The broadband operation of the phase shifter is more easily appreciated in the right part of figure 5. It is instructive to compute as well the voltage range for which the 360º phase shift is obtained. Using (6) then this voltage range is 0.12 volt. The above results show that, in principle, a graphene based broadband phase shifter is feasible, featuring the advantages of fast reconfiguration and low voltage excursion to achieve full 360º phase shift.

### 5. Summary and Conclusions

We have proposed the use of silicon graphene waveguides to implement a tunable broadband microwave photonics phase shifter based on integrated ring cavities. Numerical computation results have shown the feasibility for broadband operation over 40 GHz bandwidth and full 360º radiofrequency phase-shift with a modest voltage excursion of 0.12 volt.

### Acknowledgments

The authors wish to acknowledge the financial support given by the Research Excellency Award Program GVA PROMETEO 2013/012;

## References

1. A. K. Geim and K. S. Novoselov, “The rise of graphene,” Nat. Mater. 6, 183–191 (2007).
2. A. Vakil and N. Engheta, “Transformation optics using graphene,” Science 332, 1291–1294 (2008).
3. B. Sensale-Rodriguez et al. "Graphene for Reconfigurable THz Optoelectronics", Proceedings of the IEEE 107, 1705-1716, (2013).
4. F. Bonnacorso, Z. Sun, T. Hasan and A-C- Ferrari, "Graphene Photonics and optoelectronics", Nat. Photonics 4, 611-622, (2010).
5. Zhiwei Zheng, Chujun Zhao, Shunbin Lu, Yu Chen, Ying Li, Han Zhang, and Shuangchun Wen, "Microwave and optical saturable absorption in graphene," Opt. Express 20, 23201-23214 (2012)
6. M. Liu, X. Yin, E. Ulin-Avila, B. Geng, T. Zentgraf, L. Ju, F.Wang, and X. Zhang. “A graphene-based broadband optical modulator,” Nature 474, 64–67 (2011).
7. M. Liu, X. Yin, and X. Zhang, “Double–layer graphene optical modulator,” Nano Lett. 12, 1482–1485 (2012).
8. Z. Lu and L. Zhao, "Nanoscale electro-optic modulators based on graphene-slot waveguides", J. Opt. Soc. Am. B, 6, 1490-1496, (2012).
9. Michele Midrio, Stefano Boscolo, Michele Moresco, Marco Romagnoli, Costantino De Angelis, Andrea Locatelli, and Antonio-Daniele Capobianco, "Graphene–assisted critically–coupled optical ring modulator," Opt. Express 20, 23144-23155 (2012).
10. L. Yang et al., " Proposal for a 2×2 Optical Switch Based on Graphene-Silicon-Waveguide Microring", IEEE Photon. Tech. Lett (in press), (2013).
11. Chao Xu, Yichang Jin, Longzhi Yang, Jianyi Yang, and Xiaoqing Jiang, "Characteristics of electro-refractive modulating based on Graphene-Oxide-Silicon waveguide," Opt. Express 20, 22398-22405 (2012).
12. J. Capmany and D. Novak, “Microwave photonics combines two worlds,” Nat. Photonics 1(6), 319-330 (2007).
13. J. Yao, “Microwave photonics,” J. Lightwave Technol. 27(3), 314-335 (2009).
14. David Marpaung, Chris Roeloffzen, René Heideman, Arne Leinse, Salvador Sales, and José Capmany, "Integrated Microwave Photonics", Laser & Photonics Reviews, 7, 506–538, (2013).
15. G.W. Hanson, “Dyadic Green’s function and guided surface waves for a surface conductivity model of graphene, ” J. Appl. Phys. 103, 064302 (2008).
16. D.B. Adams and C. Madsen, " A Novel Broadband Photonic RF Phase Shifter", IEEE J. Lightwave Technol. 15, 2712-2717, (2008).
17. M. Pu, L. Liu, W. Xue, Y. Ding, L. H. Frandsen, H. Ou, K. Yvind, and J. M. Hvam, “Tunable microwave phase shifter based on silicon-on-insulator microring resonator,” IEEE Photon. Technol. Lett. 22(12), 869–871 (2010).
18. Maurizio Burla, David Marpaung, Leimeng Zhuang, Chris Roeloffzen, Muhammad Rezaul Khan, Arne Leinse, Marcel Hoekman, and René Heideman, "On-chip CMOS compatible reconfigurable optical delay line with separate carrier tuning for microwave photonic signal processing," Opt. Express 19, 21475-21484 (2011).